\documentclass[11pt,a4paper]{article}
\usepackage{amsfonts}
\usepackage{amsmath,amssymb}
\usepackage{epsfig,graphicx}

\input{tcilatex}
\begin{document}

\begin{center}
\bigskip

{\LARGE The Eightfold Way for Composite Quarks and Leptons}

{\Huge \bigskip }

\bigskip

\bigskip

\textbf{\ J.L.~Chkareuli}

\bigskip

\textit{Center for Elementary Particle Physics, ITP, Ilia State University,
0162 Tbilisi, Georgia\ \vspace{0pt}\\[0pt]
}

\textit{and} \textit{E. Andronikashvili} \textit{Institute of Physics, 0186
Tbilisi, Georgia\ }

\bigskip \bigskip \bigskip \bigskip

\bigskip \bigskip\ \ \ \ \ \ \ \ \ \ \ \ \ \ \ \ \ \ \ \ \ \ \ \ \ \ \ \ \ \
\ \ \ \ \ \ \ \ \ \ \ \ \ \ \ \ \ \ \ \ \ \ \ \ \ \ \ \ \ \ \ \ \ \ \ \ \ \
\ \ \ \ \ \ \ \ \textit{To memory of V.N. Gribov (1930-1997)} \bigskip
\bigskip\bigskip \bigskip\bigskip

\textbf{Abstract}

\bigskip
\end{center}

It is now almost clear that there is no meaningful internal symmetry higher
than the one family GUTs like as $SU(5)$, $SO(10)$, or $E(6)$ for
classification of all observed quarks and leptons. Any attempt to describe
all three quark-lepton families in the GUT framework leads to higher
symmetries with enormously extended representations which contain lots of
exotic states as well that never been detected in an experiment. This may
motivate us to continue seeking a solution in some subparticle or preon
models for quark and leptons just like as in the nineteen-sixties the
spectroscopy of hadrons had required to seek a solution in the quark model
for hadrons. At that time, there was very popular some concept invoked by
Murray Gell-Mann and called the Eightfold Way according to which all
low-lying baryons and mesons are grouped into octets. We now find that this
concept looks much more adequate when it is applied to elementary preons and
composite quarks and leptons. Remarkably, just the eight left-handed and
right-handed preons and their generic metaflavor symmetry $SU(8)$ may
determine the fundamental constituens of material world. This result for an
admissible number of preons, $N=8$, appears as a solution to the 't Hooft's
anomaly matching condition provided that (1) this condition is satisfied
separately for the $L$-preon and $R$-preon composites and (2) these
composites fill only one multiplet of some $SU(N)$ symmetry group rather
than a set of its multiplets. We next show that a partial $L$-$R$ symmetry
breaking reduces an initially emerged vectorlike $SU(8)$ theory down to the
conventional $SU(5)$ GUT with an extra local family symmetry $SU(3)_{F}$ and
three standard generations of quarks and leptons.

\bigskip \bigskip

\bigskip

\bigskip \bigskip

\bigskip

\bigskip \bigskip

\bigskip

\bigskip \bigskip

\bigskip

\bigskip

{\tiny Invited talk at the 20th International Workshop "What Comes Beyond
the Standard Model?" (9-17 July 2017, Bled, Slovenia)}

\thispagestyle{empty}\newpage

\section{Preamble}

As is well known, the Eightfold Way is the term coined by Murray Gell-Mann
in 1961 to describe a classification scheme for hadrons, that he had
devised, according to which the known\ baryons and mesons are grouped into
the eight-member families of some hadron flavor symmetry $SU(3)$ \cite{ew}.
This concept had finally led to the hypothesis of quarks locating in the
fundamental triplet of this symmetry, and consequently to a compositeness of
baryons and mesons observed. We try to show now that the eightfold way idea
looks much more adequate when it is applied to a next level of the matter
elementarity, namely, to elementary preons and composite quarks and leptons.
Remarkably, just the eight preons and their generic $SU(8)$ symmetry seem to
determine the fundamental entities of the physical world and its total
internal symmetry. Interestingly, not only the sacred number eight for
preons but also their collections in some subdivisions corresponds to the
spirit of the eightfold way that will be seen below.

In more detail, the Eightfold Way or Noble Eightfold Path \cite{nep} is a
summary of the path of Buddhist practices leading, as supposed, to a true
liberation. Keeping in mind the particle physics we propose that the eight
spoke Dharma wheel which symbolizes the Noble Eightfold Path could be
associated with eight preon fields (or superfields, in general) $P_{i}$ ($%
i=1,\ldots ,8$) being the fundamental octet of the basic\ flavor symmetry $%
SU(8)$. They may carry out the eight fundamental quantum numbers which has
been detected so far. These numbers are related to the weak isospin, color
and families of quarks and leptons. Accordingly, we will refer to these
preons as a collection of "isons" $P_{w}$ ($w=1,2$), "chromons" $P_{c}$ ($%
c=1,2,3$) and "famons" $P_{f}$ ($f=1,2,3$). Surprisingly, the Noble
Eightfold Path is also originally divided into three similar basic
divisions. They are: (1) The Insight consisting of the Right view and the
Right resolve, (2) The Moral virtue consisting of the Right speech, the
Right action and the Right livelihood and (3) The Meditation consisting of
the Right effort, the Right mindfulness and the Right Concentration. An
analogous decomposition of the "sacred number eight", $8=2+3+3$, which
appears in the expected breakdown of the generic preon $SU(8)$ symmetry%
\begin{equation}
SU(8)\rightarrow SU(2)_{W}\times SU(3)_{C}\times SU(3)_{F}\text{ ,}
\label{8}
\end{equation}%
looks indeed rather impressive.

In principle, it is not necessary to generically relate the Eightfold Way
concept to preons and composite quarks and leptons. First of all, it is
related to the eight fundamental quantum charges of particle physics
presently observed. They correspond in fact to the two weak isospin
orientations, the three types of colors and the three species of
quark-lepton families, all of which may be accommodated in the unified $%
SU(8) $ theory. Their carriers could be or could not be the elementary
preons, though the preon model composing the observed quark and leptons at
appropriate distances seems to reflect this concept in the most transparent
way.

We find, resurrecting to an extent the old Eightfold Way idea in an
initially $L$-$R$ symmetric and $SU(N)$ invariant physical world, that just
the eight left-handed and right-handed preons and their basic flavor
symmetry $SU(8)$ appear as a solution to the 't Hooft's anomaly matching
condition \cite{t} providing the chiral symmetry preservation at all
distances involved and, therefore, masslessness of emerged composite
fermions. We show that this happens if (1) this condition is satisfied
separately for the $L$-preon and $R$-preon composites and (2) each of these
two series of composites fill only one irreducible representation of the
starting $SU(N)$ symmetry group rather than a set of its representations.
However, such an emerged $L$-$R$ symmetric $SU(8)$ theory, though is chiral
with respect to preons, certainly appears vectorlike for the identical $L$%
-preon and $R$-preon composite multiplets involved. This means that, while
preons are left massless being protected by their own metacolors, the
composites being metacolor singlets will pair up and acquire heavy Dirac
masses. We next show how an appropriate $L$-$R$ symmetry violation reduces
the metaflavor $SU(8)$ theory down to one of its chiral remnants being of
significant physical interest. Particularly, this violation implies that,
while there still remains the starting chiral symmetry $SU(8)_{L}$ for the
left-handed preons and their composites, for the right-handed states\ we may
only have the broken chiral symmetry $[SU(5)\times SU(3)]_{R}$. Therefore,
whereas nothing really happens with the left-handed preon composites still
filling the total multiplet of the $SU(8)$, the right-handed preon
composites will form only some particular submultiplets in it. As a result,
we eventually come to the conventional $SU(5)$ GUT with an extra local
family symmetry $SU(3)_{F}$ and three standard generations of quarks and
leptons. Moreover, the theory has the universal gauge coupling constant
running down from the $SU(8)$ unification scale, and also predicts some
extra heavy $SU(5)\times SU(3)_{F}$ multiplets located at the scales from $%
O(1)$ $TeV$ up to the Planck mass that may appear of actual experimental
interest. For simplicity, we largely work in an ordinary spacetime
framework, though extension to the conventional $N=1$ supersymmetry with
preons and composites treated as standard scalar superfields could generally
be made. All these issues are successively considered in the subsequent
sections 2-7, and in the final section 8 we present our conclusion.

Some attempt to classify quark-lepton families in the framework of the $%
SU(8) $ GUT with composite quarks and leptons had been made quite a long ago 
\cite{jp}, though with some special requirements which presently seem not
necessary or could be in principle derived rather than postulated. Since
then also many other things became better understood, especially the fact
that the chiral family symmetry subgroup $SU(3)_{F}$ of the $SU(8)$, taken
by its own, was turned out to be rather successful in description of
quark-lepton generations. At the same time, there have not yet appeared, as
mentioned above, any other meaningful internal symmetry for an appropriate
classification of all the observed quarks and leptons. All that motivates us
to address this essential problem once again.

\section{Preons - mataflavors and metacolors}

We start formulating a few key elements of preon models (for some
significant references, see \cite{moh, ds}), partially refining some issues
given in our old paper \cite{jp}.

\begin{itemize}
\item We propose that there is an exact $L$-$R$ symmetry at small distances
where $N$ elementary massless left-handed and right-handed preons, $P_{iL}$
and\ $P_{iR}$ \ \ $(i=1,\ldots ,N)$, possess a local metaflavor symmetry $%
SU(N)_{MF}$ including the known physical charges, such as weak isospin,
color, and family number. The preons, both $P_{iL}$ and $P_{iR}$, transform
under its fundamental representation.

\item The preons also possess a local metacolor symmetry $%
G_{MC}=G_{MC}^{L}\times G_{MC}^{R}$ with $n$ metacolors\ ($n$ is odd) which
bind preons into composites - quarks, leptons and other states. In contrast
to their common metaflavors, the left-right and left-handed preons, $%
P_{iL}^{\alpha }$ and \ $P_{iR}^{\alpha ^{\prime }}$, have different
metacolors, where ${\alpha }$ and ${{\alpha }^{\prime }}\ $are indices of
the corresponding metacolor subgroups $G_{MC}^{L}$ (${\alpha =1,...,n}$) and 
$G_{MC}^{R}${\ (}${{\alpha }^{\prime }=1,...,n}$), respectively. As a
consequence, there are two types of composites at large distances being
composed from them separately with a radius of compositeness, $R_{MC}$ $\sim
1/\Lambda _{MC}$, where $\Lambda _{MC}$ corresponds to the scale of the
preon confinement for the asymptotically free (or infrared divergent)
metacolor symmetries. Due to the proposed $L$-$R$ symmetry, the metacolor
symmetry groups $G_{MC}^{L}$ and $G_{MC}^{R}$ are taken identical with the
similar scales for both of sets of preons. If one also proposes that the
preon metacolor symmetry $G_{MC}$ is generically anomaly-free for any matter
content involved, one comes to an input chiral orthogonal symmetry of the
type%
\begin{equation}
G_{MC}=SO(n)_{MC}^{L}\times SO(n)_{MC}^{R}\text{, \ }n=3,5,...  \label{lrmc}
\end{equation}%
for the $n$-preon configurations of composites. For reasons of economy, it
is usually proposed that the emerged fermion composites have the minimal $3$%
-preon configuration. Obviously, the preon condensate $\left\langle 
\overline{P_{L}}P_{R}\right\rangle $ which could cause the $\Lambda _{MC}$
order masses for composites is principally impossible in the two-metacolor
model (\ref{lrmc}). This is in sharp contrast to an ordinary QCD case where
the left-handed and right-handed quarks forms the $\left\langle \overline{%
q_{L}}q_{R}\right\rangle $ condensate thus leading to the $\Lambda _{C}$
order masses ($\Lambda _{C}$ $\sim (0.1\div 1)GeV$) for composite mesons and
baryons.

\item Apart from the local symmetries, metacolors and metaflavors, the
preons $P_{iL}^{\alpha }$ and \ $P_{iR}^{\alpha ^{\prime }}$ possess an
accompanying chiral global symmetry 
\begin{equation}
K(N)=SU(N)_{L}\times SU(N)_{R}  \label{ch}
\end{equation}%
being unbroken at the small distances. This symmetry is typically considered
in the limit when the $SU(N)_{MF}$ gauge metaflavor interactions are
switched off. Indeed, these interactions are too weak to influence the bound
state spectrum. We omitted above the Abelian chiral $U(1)_{L,R}$ symmetries
in $K(N)$ since the corresponding currents have Adler-Bell-Jackiw anomalies
in the triangle graph where they couple to two metagluons \cite{th}. In
fact, their divergences for massless preons are given by 
\begin{equation}
\partial _{\mu }J_{L,R}^{\mu }=n\frac{g_{L,R}^{2}}{16\pi ^{2}}G_{L,R}^{\mu
\nu }G_{L,R}^{\rho \sigma }\epsilon _{\mu \nu \rho \sigma }  \label{div}
\end{equation}%
where $G_{L,R}^{\mu \nu }$ are the metagluon field strengths for the $%
SO(n)_{MC}^{L,R}$ metacolors, respectively, while $g_{L,R}$ are the
appropriate gauge coupling constants. Thereby, the chiral symmetries $%
U(1)_{L,R}$, which would present the conserved chiral hypercharges in the
classical Lagrangian with massless preons, appear broken by the quantum
corrections (\ref{div}) that make us to leave only the non-Abelian chiral
symmetry (\ref{ch}) in the theory. Nevertheless, one could presumably still
use these symmetries at the small distances, $r\ll $ $R_{MC}$, where the
corrections (\ref{div}) may become unessential due to asymptotic freedom in
the metacolor theory considered. We will refer to this regime as the valent
preon approximation in which one may individually recognize each preon no
matter it is free or bound in a composite fermion. Therefore, the chiral
preon numbers or hypercharges $Y_{L,R}$ related to the symmetries $%
U(1)_{L,R} $ may be considered in this approximation\ as the almost
conserved classical charges according to which the preon and composite
states are allowed to be classified.

\item The fact that the left-handed and right-handed preons do not form the $%
\left\langle \overline{P_{L}}P_{R}\right\rangle $ condensate may be
generally considered as a necessary but not yet a sufficient condition for
masslessness of composites. The genuine massless fermion composites are
presumably only those which preserve chiral symmetry of preons (\ref{ch}) at
large distances that is controlled by the 't Hooft's anomaly matching (AM)
condition \cite{t}. For reasons of simplicity, we do not consider below
boson composites, the effective scalar or vector fields. Generally, they
being no protected by any symmetry will become very heavy (with masses of
the order of the compositeness scale $\Lambda _{MC}$) and decouple from a
low-lying particle spectrum.
\end{itemize}

\section{AM conditions for N metaflavors}

The AM condition \cite{t}\ states in general that triangle anomalies related
to $N$\ massless elementary preons, both left-handed and right-handed, have
to match those for massless fermions (including quarks and leptons) being
composed by the metacolor forces arranged by the proposed local symmetry $%
SO(n)_{MC}^{L}\times SO(n)_{MC}^{R}$. Based on the starting $L$-$R$ symmetry
in our model we will require, in some contrast to the original AM condition 
\cite{t}, that fermions composed from the left-handed and right-handed
preons with their own metacolors, $SO(n)_{MC}^{L}$ and $SO(n)_{MC}^{R}$,
have to satisfy the AM condition separately, whereas the metaflavor triangle
anomalies of the $L$-preon and $R$-preon composites may in general
compensate each other. Therefore, in our $L$-$R$ symmetric preon model one
does not need to specially introduce elementary metacolor singlet fermions,
called the "spectator fermions" \cite{t}, to cancel these anomalies both at
the small and large distances.

The AM condition puts in general powerful constraints on the classification
of massless composite fermions with respect to the underlying local
metaflavor symmetry $SU(N)_{MF}$ or some of its subgroups, depending on the
extent to which the accompanying global chiral symmetry (\ref{ch}) of preons
remains at large distances. In one way or another, the AM condition 
\begin{equation}
\sum_{r}i_{r}a(r)=na(N)  \label{am}
\end{equation}%
for preons (the right side) and composite fermions (the left side) should be
satisfied. Here $a(N)$ and $a(r)$ are the group coefficients of triangle
anomalies related to the groups $SU(N)_{L}$ or $SU(N)_{R}$ in (\ref{ch})
whose coefficients are calculated in an ordinary way, \ \ \ \ 
\begin{equation}
a(r)d^{ABC}=Tr(\{T^{A}T^{B}\}T^{C})_{r}\text{ , \ }Tr(T^{A}T^{B})=\frac{1}{2}%
\delta ^{AB}  \label{a}
\end{equation}%
where $T^{A}$ ($A,B,C=1,\ldots ,N^{2}-1$) are the $SU(N)$ generators taken
in the corresponding representation $r$. The $a(N)$ corresponds a
fundamental representation and is trivially equal to $\pm 1$ (for
left-handed and right-handed preons, respectively), while $a(r)$ is related
to a representation $r$ for massless composite fermions. The values of the
factors $i_{r}$ give a number of times the representation $r$ appears in a
spectrum of composite fermions and are taken positive for the left-handed
states and negative for the right-handed ones.

The anomaly coefficients for composites $a(r)$ contain an explicit
dependence on the number of preons $N$, due to which one could try to find
this number from the AM condition. However, in general, there are too many
solutions to the condition (\ref{am}) for any value of $N$. Nevertheless,
for some special, though natural, requirements an actual solution may only
appear for $N=8$, as we will see below.

Indeed, to strengthen the AM condition one could think that it would more
appropriate to have all composite quarks and leptons in a single
representation of the unified symmetry group rather than in some set of its
representations. This, though would not largely influence the gauge sector
of the unified theory, could make its Yukawa sector much less arbitrary.
Apart from that, the composites belonging to different representations would
have in general different preon numbers that could look rather unnatural.
Let us propose for the moment that we only have the minimal three-preon
fermion composites formed by the metacolor forces which correspond to the $%
SO(3)_{MC}^{L}\times SO(3)_{MC}^{R}$ symmetry case in (\ref{lrmc}). We will,
therefore, require that only some single representation $r_{0}$ for massless
three-preon states has to satisfy the AM condition that simply gives in (\ref%
{am}) 
\begin{equation}
a(r_{0})=3\text{ }  \label{am1}
\end{equation}%
individually for $L$-preon and $R$-preon composites.

Now, calculating the anomaly coefficients for all possible three $L$-preon
and three $R$-preon composites one respectively has (see also \cite{jp, bar})%
\begin{eqnarray}
&&\Psi _{\{ijk\}L,R}\text{ , }N^{2}/2+9N/2+9,  \notag \\
&&\Psi _{\lbrack ijk]L,R}\text{ , \ }N^{2}/2-9N/2+9,  \notag \\
&&\Psi _{\{[ij]k\}L,R}\text{, \ }N^{2}-9,  \notag \\
&&\Psi _{\{jk\}L,R}^{i}\text{ , \ }N^{2}/2+7N/2-1,  \notag \\
&&\Psi _{\lbrack jk]L,R}^{i}\text{ , \ }N^{2}/2-7N/2-1  \label{tens}
\end{eqnarray}%
with all appropriate $SU(N)_{L,R}$\ representations listed (anomaly
coefficients for right-handed composites have to be taken with an opposite
sign). Putting then each of the above anomaly coefficients in (\ref{tens})
into the AM condition (\ref{am1}) one can readily find that there is a
solution with an integer $N$ only for the last tensors $\Psi _{\lbrack
jk]L,R}^{i}$ , and this is in fact the unique "eightfold" solution 
\begin{equation}
N^{2}/2-7N/2-1=3,\text{ }N=8\text{ .}  \label{8-1}
\end{equation}%
Remarkably, the same solution $N=8$ appears independently, if one requires
that some $SU(N)$ symmetry has to possess the right $SU(5)$ GUT assignment 
\cite{moh} for the observed quark-lepton families in order to be in fact in
accordance with observations. This means that one of its $3$-index
representations has to contain an equal numbers of the $SU(5)$ anti-quintets 
$\overline{5}^{k}$ and decuplets $10_{[k,l]}$ ($k,l$ are the $SU(5)$
indices). Indeed, decomposing $SU(N)$ into $SU(5)\times $ $SU(N-5)$ one find
that this equality exists only for the representation $\Psi _{\lbrack
jk]}^{i}$ in (\ref{tens}) that reads as%
\begin{equation}
(N-5)(N-6)/2=N-5,\text{ }N=8\text{ }  \label{8-2}
\end{equation}%
thus leading again to the "eightfold" $SU(8)$ metaflavor symmetry.

Let us note that, apart the minimal 3-preon states, there are also possible
some alternative higher preon configurations for \ composite quarks and
leptons. Moreover, the $SO(3)_{MC}^{L,R}$ metacolors providing the
three-preon structure of composite quarks and leptons may appear
insufficient for the preon confinement, unless one invokes some special
strong coupling regime \cite{wil}. For the asymptotically free $SO(n)$
metacolor, one must generally require $n>2+2N/11$, due to which the
composite quarks and leptons have to appear at least as the five-preon
states. Checking generally all possible $n$-index representations of $SU(N)$
we find that the AM condition only works for some combination of its
"traced" tensors $\Psi _{\lbrack jk]L,R}^{i}$ and $\Psi _{iL,R}$ obtained
after taking traces out of the proper $n$-index tensors $\Psi _{\lbrack
jk...]L,R}^{i...}$. This eventually leads to the equation generalizing the
above anomaly matching condition (\ref{8-1})%
\begin{equation}
N^{2}/2-7N/2-1+p=n  \label{np}
\end{equation}%
where $p$ is a number of the traced fundamental multiplets $\Psi _{iL,R}$
for composites. One can see that there appear some reasonable solutions only
for $n-p=3$ and, therefore, one has again solutions for the "eightfold"
metaflavor symmetry $SU(8)$.

Apart from the AM condition (\ref{am}) there would be in general another
kind of constraint on composite models which has been also proposed in \cite%
{t}. This constraint requires the anomaly matching for preons and
composites, even if some of introduced $N$ preons become successively
heavier than the scale of compositeness and consequently decouple from the
entire theory. As a result, the AM condition should work for any number of
preons remained massless (thus basically being independent of $N$) that
could make generally classification of composite fermions quite arbitrary.
Fortunately, such an extra constraint is not applicable to our chiral
two-metacolor model where the Dirac masses for preons are not possible by
definition, whereas the Majorana masses would mean breaking of the input
local metaflavor $SU(N)_{MF}$ symmetry\footnote{%
Apart from that, it has been generally argued \cite{moh} that the
nonperturbative effects may not be analytic in the preon mass so that for
the large and small preon masses the theories may be quite different, thus
avoiding this additional constraint.}.

Most importantly, the orthogonal symmetry for metacolor (\ref{lrmc}) allows
to consider more possible composite configurations than it is in the case of
an unitary metacolor symmetry, as in the conventional $SU(3)$ color for QCD.
The above strengthening of the AM condition, according to which the
composites only fill a single multiplet of the metaflavor $SU(N)_{MF}$
symmetry group, has unambiguously led us to the composite multiplets $\Psi
_{\lbrack jk]L,R}^{i}$ having the same classical $U(1)_{L,R}$ fermion
numbers or hypercharges $Y_{L,R}$ as the preons themselves. We argued in the
previous section that these hypercharges may be considered in the valent
preon approximation\ as the almost conserved classical charges according to
which the preon and composite states could be classified. With all that in
mind, one could assume that there may work some extra selection rule
according to which only composites satisfying the condition%
\begin{equation}
Y_{L,R}(preons)=Y_{L,R}(composites)  \label{yy}
\end{equation}%
appear in physical spectrum in the orthogonal left-right metacolor case.

We can directly see that the condition (\ref{yy}) trivially works for the
simplest composite states which could be constructed out of a single preon $%
P_{iL}^{\alpha }$ or\ $P_{iR}^{\alpha ^{\prime }}$, whose metacolor charge
is screened by the metagluon fields $A_{L\mu }^{\alpha }$ and $A_{R\mu
}^{\alpha ^{\prime }}$ of $SO(n)_{MC}^{L}$ and $SO(n)_{MC}^{R}$,
respectively. These composites will also satisfy the general AM condition (%
\ref{am}) provided that one admits the $n$ left-handed and right-handed
fundamental composite multiplets of the $SU(N)_{MF}$ to participate ($%
i_{N}=n $). In our $L$-$R$ symmetric model, however, such massless
composites will necessarily pair up, thus becoming very massive and
decoupling from the low-lying particle spectrum, no matter the starting $L$-$%
R$ symmetry becomes later broken or not. This in sharp contrast to the
models with the orthogonal metacolor group $SO(n)$ for the single chirality
preons \cite{bar}, where such massless composite generally appear to be in
contradiction with observations. Moreover, in this case the composite
multi-preon states for quarks and leptons seem hardly to be stable, since
they could freely dissociate into the screened preon states.

One could wonder why the condition (\ref{yy}) does not work in the familiar
QCD case with elementary quarks and composite baryons. The point is that,
despite some conceptual similarity, QCD is the principally different theory.
The first and immediate is that the unitary color $SU(3)_{C}$, in contrast
to the orthogonal ones, allows by definition no other quark number for
baryons but $Y_{B}=3Y_{q}$. The most important aspect of this difference is,
however, that the color symmetry $SU(3)_{C}$ is vectorlike due to which
chiral symmetry in QCD is broken by quark-antiquark condensates with the
corresponding zero-mass Goldstone bosons (pions, kaons etc.) providing the
singularity of the three-point function. As a consequence, the AM condition
implies in this case that dynamics requires a spontaneous breakdown of
chiral symmetry rather than an existence of massless composite fermions, as
happens in the orthogonal metacolor case discussed above.

We find below in section 5 that, though the proposed condition (\ref{yy})
looks rather trivial in the $L$-$R$ symmetry phase of the theory, it may
become rather significant when this symmetry becomes spontaneously broken.

\section{Composites - the L-R symmetry phase}

So, we have at small distances the preons given by the Weil fields 
\begin{equation}
P_{iL}^{\alpha }\text{ , \ }P_{iR}^{\alpha ^{\prime }}\text{ \ \ \ \ }%
(i=1,\ldots ,8;\text{ }{\alpha }=1,2,3;\text{ }{\alpha }^{\prime }=1,2,3)
\label{88}
\end{equation}%
belonging to the fundamental octet of the local metaflavor symmetry $%
SU(8)_{MF}$ and to triplets of the metacolor symmetry $SO(3)_{MC}^{L}\times
SO(3)_{MC}^{R}$ which are local, and there is also the accompanying global
chiral symmetry%
\begin{equation}
K(8)=SU(8)_{L}\times SU(8)_{R}  \label{881}
\end{equation}%
of the eight preon species (\ref{88}). At large distances, on the other
hand, we have composites which are, respectively, in the left-handed and
right-handed multiplets of the $SU(8)_{MF}$ 
\begin{equation}
\Psi _{\lbrack jk]L}^{i}(216)\text{ , }\Psi _{\lbrack jk]R}^{i}(216)\text{ ,}
\label{216}
\end{equation}%
where their dimensions are explicitly indicated. The chiral symmetry (\ref%
{881}), according to the AM condition taken, remains at large distances. Due
to a total $L$-$R$ symmetry of preons and composites the metaflavor triangle
anomalies at both small and large distances appears automatically
compensated. Decomposing the $SU(8)_{MF}$\ composite multiplets (\ref{216})
into the $SU(5)\times SU(3)_{F}$ components one has%
\begin{equation}
216_{L,R}=\left[ (\overline{5}+10,\text{ }\overline{3}%
)+(45,1)+(5,8)+(24,3)+(1,3)+(1,\overline{6})\right] _{L,R}  \label{216'}
\end{equation}%
where the first term for the left-handed composites, $(\overline{5}+10,$ $%
\overline{3})_{L}$, could be associated with the standard $SU(5)$ GUT\
assignment for quarks and leptons \cite{moh} extended by some family
symmetry $SU(3)_{F}$, while the other multiplets are somewhat exotic and,
hopefully, could be made heavy to decouple them from an observed low-lying
particle spectrum.

The determination of an explicit form of wave functions for the composite
states (\ref{216}) is a complicated dynamical problem related to the yet
unknown dynamics of the preon confinement. We propose that some basic
feature of these composites are simply given by an expression%
\begin{equation}
\Psi _{\lbrack jk]L}^{i}(x)\propto \epsilon _{\alpha \beta \gamma }\left( 
\overline{P}_{L}^{\alpha i}\gamma _{\mu }P_{L[j}^{\beta }\right) \gamma
^{\mu }P_{k]L}^{\gamma }(x)\text{ \ }  \label{wf}
\end{equation}%
where indices $\alpha $, $\beta $, $\gamma $ belong to the metacolor
symmetry $SO(3)_{MC}^{L}$. In the valent preon approximation, the preon
current (\ref{wf}) corresponds to a bound state of three left-handed preons
with zero mass ($p^{2}=(p_{1}+p_{2}+p_{3})^{2}=0$) being formed by massless
preons ($p_{1}^{2}=p_{2}^{2}=p_{3}^{2}=0$) which are moving in a common
direction. It is then clear that a state with a spin of $1/2$ (and a
helicity $-1/2$) can be only obtained by assembling two preons and one
antipreon into a quark or lepton. In a similar way one can construct the
preon current $\Psi _{\lbrack jk]R}^{i}$ which corresponds to a multiplet of
states again with a spin of $1/2$ (but a helicity $+1/2$) composed from
right-handed preons. This is simply achieved by making the proper
replacements in (\ref{wf}) leading to the composite states 
\begin{equation}
\Psi _{\lbrack jk]R}^{i}(x)\propto \epsilon _{\alpha ^{\prime }\beta
^{\prime }\gamma ^{\prime }}\left( \overline{P}_{R}^{\alpha ^{\prime
}i}\gamma _{\mu }P_{R[j}^{\beta ^{\prime }}\right) \gamma ^{\mu
}P_{k]R}^{\gamma ^{\prime }}(x)  \label{wf1}
\end{equation}%
where indices $\alpha ^{\prime }$, $\beta ^{\prime }$, $\gamma ^{\prime }$
belong now to the metacolor symmetry $SO(3)_{MC}^{R}$. For the simplest
composite states which can be constructed out of a single preon $%
P_{iL}^{\alpha }$ or\ $P_{iR}^{\alpha ^{\prime }}$ , whose metacolor charge
is screened by the metagluon fields $A_{L\mu }^{\alpha }$ and $A_{R\mu
}^{\alpha ^{\prime }}$ of $SO(3)_{MC}^{L}$ and $SO(3)_{MC}^{R}$, the wave
functions may be written as%
\begin{equation}
\Psi _{iL}(x)\propto A_{L\mu }^{\alpha }\gamma ^{\mu }P_{iL}^{\alpha }(x)%
\text{ , \ }\Psi _{iR}(x)\propto A_{R\mu }^{\alpha ^{\prime }}\gamma ^{\mu
}P_{iR}^{\alpha ^{\prime }}(x)\text{ ,}  \label{wf2}
\end{equation}%
respectively.

Let us remark in conclusion that the whole theory so far considered, though
looks chiral with respect to preons (\ref{88}), is certainly vectorlike for
composites (\ref{216}). This means that, while preons are left massless
being protected by their own metacolors, all the $L$-preon and $R$-preon
composites being metacolor singlets will pair up due to some quantum
gravitational transitions and, therefore, acquire Dirac masses. We find
below that due to closeness of the compositeness scale $\Lambda _{MC}$ to
the Planck scale $M_{Pl}$ the masses of all composites appear very heavy
that has nothing in common with reality. It is rather clear that such a
theory is meaningless unless the proposed $L$-$R$ symmetry is somehow broken
that seems to be in essence a basic point in our model. One could expect
that such breaking may eventually exclude the right-handed submultiplet $(%
\overline{5}+10,$ $\overline{3})_{R}$ in the composite spectrum (\ref{216'}%
), while leaving there its left-handed counterpart, $(\overline{5}+10,$ $%
\overline{3})_{L}$, which can be then uniquely associated with the observed
three families of ordinary quarks and leptons.

\section{Composites - partially broken L-R symmetry}

We propose that there is a partial breaking of the chiral symmetry (\ref{881}%
) in the right-handed preon sector of the type 
\begin{equation}
K(8)\rightarrow SU(8)_{L}\times \lbrack SU(5)\times SU(3)]_{R}  \label{l-r}
\end{equation}%
being considered in the zero limit for the $SU(8)_{MF}$ metaflavor gauge
coupling constant. For convenience, we consider some supersymmetric model
for preons and composites where this breaking\ may be presumably caused by
an asymmetric preon condensation%
\begin{equation}
\epsilon _{\alpha \beta \gamma }\left\langle P_{iL}^{\alpha }P_{jL}^{\beta
}P_{kL}^{\gamma }\right\rangle =0\text{ , \ }\epsilon _{\alpha ^{\prime
}\beta ^{\prime }\gamma ^{\prime }}\left\langle P_{iR}^{\alpha ^{\prime
}}P_{jR}^{\beta ^{\prime }}P_{kR}^{\gamma ^{\prime }}\right\rangle =\delta
_{i}^{a}\delta _{j}^{b}\delta _{k}^{c}\epsilon _{abc}\Lambda _{MC}^{4}\text{
\ \ }  \label{con}
\end{equation}%
emerging for preon superfields with their fermion and scalar field
components involved. Here antisymmetric third-rank tensors $\epsilon
_{\alpha \beta \gamma }$ and $\epsilon _{\alpha ^{\prime }\beta ^{\prime
}\gamma ^{\prime }}$ belong to the metacolor symmetries $SO(3)_{MC}^{L}$ and 
$SO(3)_{MC}^{R}$, respectively, while $\epsilon _{abc}$ ($a,b,c=1,2,3$) to
the symmetry $SU(3)_{R}$. Remarkably, the breaking (\ref{l-r}) is only
possible when the number of metacolors $n$ is equal $3$ or $5$, as is
actually implied in our model. For the minimal case, $n=3$, the vacuum
configurations (\ref{con}) could spontaneously appear in some $L$-$R$
symmetric model with the properly arranged high-dimensional preon
interactions\footnote{%
This $L$-$R$ symmetry breaking model looks somewhat similar to the
well-known multi-fermion interaction schemes used in the other contexts for
chiral symmetry breaking \cite{io} or spontaneous Lorentz violation \cite{jb}%
.} 
\begin{eqnarray}
&&\sum_{n=1}^{\infty }\text{ }{\LARGE \{}\mathrm{G}_{n}^{LL}\left[ \left( 
\overline{P}_{L}\overline{P}_{L}\overline{P}_{L}\right) \left(
P_{L}P_{L}P_{L}\right) \right] ^{n}+\mathrm{G}_{n}^{RR}\left[ \left( 
\overline{P}_{R}\overline{P}_{R}\overline{P}_{R}\right) \left(
P_{R}P_{R}P_{R}\right) \right] ^{n}  \notag \\
&&+\mathrm{G}_{n}^{LR}\left[ \left( \overline{P}_{L}\overline{P}_{L}%
\overline{P}_{L}\right) \left( P_{R}P_{R}P_{R}\right) \right] ^{n}+\mathrm{G}%
_{n}^{RL}[\left( \overline{P}_{R}\overline{P}_{R}\overline{P}_{L}\right)
\left( P_{L}P_{L}P_{L}\right) ]^{n}{\LARGE \}}  \label{4f}
\end{eqnarray}%
with coupling constants satisfying the conditions $\mathrm{G}_{n}^{LL}=%
\mathrm{G}_{n}^{RR}$ and\ $\mathrm{G}_{n}^{LR}=\mathrm{G}_{n}^{RL}$. This
model is evidently non-renormalizable and can be only considered as an
effective theory valid at sufficiently low energies. The dimensionful
couplings \textrm{G}$_{n}$ are proportional to appropriate powers of some UV
cutoff $\Lambda $ which in our case can be ultimately related to the preon
confinement energy scale $\Lambda _{MC}$, $\mathrm{G}_{n}\sim \Lambda
_{MC}^{4-8n}$. For some natural choice of these coupling constants one may
come to the asymmetric solution (\ref{con}).

A more conventional way of getting the $L$-$R$ asymmetry may follow from the
symmetric scalar field potential \cite{moh} 
\begin{equation}
U=M^{2}(\Phi _{L}^{2}+\Phi _{R}^{2})+h(\Phi _{L}^{2}+\Phi
_{R}^{2})^{2}+h^{\prime }\Phi _{L}^{2}\Phi _{R}^{2}+P(\Phi _{L},\Phi _{R})
\label{u}
\end{equation}%
containing two elementary third-rank antisymmetric scalar fields, $\Phi
_{L}^{[ijk]}$ and $\Phi _{R}^{[ijk]}$, interacting with $L$- and $R$-preons,
respectively. For some natural area of the parameters in the potential, $%
M^{2}<0$ and $h,h^{\prime }>0$, and properly chosen couplings for scalars $%
\Phi _{L}^{[ijk]}$ and $\Phi _{R}^{[ijk]}$ in the polynomial $P(\Phi
_{L},\Phi _{R})$ they may readily develop the totally asymmetric VEV
configuration%
\begin{equation}
\left\langle \Phi _{L}^{[ijk]}\right\rangle =0\text{ , \ }\left\langle \Phi
_{R}^{[ijk]}\right\rangle =\delta _{a}^{i}\delta _{b}^{j}\delta
_{c}^{k}\epsilon ^{abc}M_{LR}\text{ \ \ \ }(a,b,c=1,2,3)  \label{ph}
\end{equation}%
where the mass $M_{LR}$ corresponds to the $L$-$R$ symmetry breaking scale
and indices $a,b,c$ belong to the $SU(3)_{R}$. Due to these VEVs, the higher
dimension terms in the effective superpotential induced generally by gravity 
\begin{equation}
\frac{\mathrm{G}_{L}}{M_{Pl}}(P_{iL}P_{jL}P_{kL})\Phi _{L}^{[ijk]}+\frac{%
\mathrm{G}_{R}}{M_{Pl}}(P_{iR}P_{jR}P_{kR})\Phi _{R}^{[ijk]}  \label{sup}
\end{equation}%
($\mathrm{G}_{L,R}$ are dimensionless coupling constants) will change the AM
conditions for right-handed states leaving those for the left-handed ones
intact. This modification is related to an appearance of the new Yukawa
interaction for preons 
\begin{equation}
\mathrm{G}_{R}^{\prime }\epsilon
^{abc}(P_{aR}^{(f)}CP_{bR}^{(f)})P_{cR}^{(s)}\text{ , \ }\mathrm{G}%
_{R}^{\prime }=\mathrm{G}_{R}\frac{M_{LR}}{M_{Pl}}\text{ }  \label{n}
\end{equation}%
where $P_{iR}^{(f,s)}$ are, respectively, the fermion and scalar field
components of the right-handed preon superfield $P_{iR}$. This interaction
will give some extra radiative corrections to the triangle graphs with
circulating "family" preons $P_{aR}^{(f)}$ ($a=1,2,3$) and their composites.
As a result, the triangle anomalies corresponding to all generators of the $%
SU(8)_{R}$, besides those of the $[SU(5)\times SU(3)]_{R}$, are left
uncompensated, that causes the proper decreasing of the chiral symmetry,
just as is indicated in (\ref{l-r}).

Eventually, while there still remains the starting chiral symmetry $%
SU(8)_{L} $ for the left-handed preons and their composites, for the
right-handed states\ we only have the broken symmetry given in (\ref{l-r}).
Therefore, whereas nothing changes for the $L$-preon composites filling the
total multiplet $216_{L}$ in (\ref{216'}), the $R$-preons will only compose
some particular submultiplets in $216_{R}$ (\ref{216'}). In general, these
submultiplets may not include the three right-handed quark-lepton families $(%
\overline{5}+10,$ $\overline{3})_{R}$. We can simply postulate it as some
possible ansatz being allowed by the different chiral symmetries in the $L$%
-preon and $R$-preon sectors in the $L$-$R$ symmetry broken phase.
Nonetheless, it would be interesting to argue using the preon number
matching condition (\ref{yy}) which we discussed in section 3. Note first
that the $U(1)_{R}$ symmetry in the right-handed sector reduces after the $L$%
-$R$ symmetry breaking (\ref{con}, \ref{ph}) to%
\begin{equation}
U(1)_{R}\rightarrow U(1)_{R}^{(5)}\times Z(3)_{R}^{(3)}  \label{pr}
\end{equation}%
while the $U(1)_{L}$ symmetry is left intact. Here, $U(1)_{R}^{(5)}$ and $%
Z(3)_{R}^{(3)}$ stand for the survived continuous and discrete symmetries of
quintet preons\ $P_{sR}$ ($s=1,...,5$) of $SU(5)_{R}$ and triplet preons $%
P_{aR}$ ($a=1,2,3$) of $SU(3)_{R}$, respectively, which are thereby
separated. Namely, the $R$-preon hypercharge group in the broken $L$-$R$
symmetry phase is given by the product (\ref{pr}) rather than the universal $%
U(1)_{R}$ for all eight preons, as was in its unbroken phase. Now, if we
require the preon number matching for preons and composites the states
collected in $(\overline{5}+10,$ $\overline{3})_{R}$ will never appear in
physical spectrum. Indeed, as one can easily check, both the $U(1)_{R}^{(5)}$
hypercharge and discrete $Z(3)_{R}^{(3)}$ symmetry values for these states
are quite different from those for the preons $P_{sR}$ and $P_{aR}$,
respectively. At the same time, all other composite submultiplets in $%
216_{R} $ (\ref{216'}) readily match the both symmetry values for preons.

One way or another, the simplest combination of the $216_{R}$ submultiplets
which may simultaneously satisfy the AM conditions for the $[SU(5)\times
SU(3)]_{R}$ symmetry, as well as the above preon number matching condition
is in fact given by the collection 
\begin{equation}
(45,1)_{R}+(5,8+1)_{R}+3(1,3)_{R}  \label{45}
\end{equation}%
where the submultiplet $(1,3)_{R}$ has to appear three times in (\ref{45})\
in order to appropriately restore the anomaly coefficient balance for the $R$%
-preon composites. Of course, this collection of states can also appear by
its own without any reference to the preon number matching condition that we
have used above as some merely heuristic argument.

\section{Physical sector - quarks and leptons}

We can see that after chiral symmetry breaking in the right-handed preon
sector through the VEVs (\ref{con}) or (\ref{ph}) the starting metaflavor
symmetry $SU(8)_{MF}$ at large distances is reduced to the product of the
standard $SU(5)$ GUT and chiral family symmetry $SU(3)_{F}$ 
\begin{equation}
SU(8)_{MF}\rightarrow SU(5)\times SU(3)_{F}\text{ }  \label{gf}
\end{equation}%
presumably with the equal gauge coupling constants $g_{5}$ and $g_{3F}$ at
the unification scale. This is in essence the chiral remnant of the
initially emerged vectorlike $SU(8)_{MF}$ symmetry. The massless composite
fermions, due to pairing up of the similar $L$-preon and $R$-preon
composites and decoupling them from a low-energy spectrum, are given now by
the collection of the $SU(5)\times SU(3)_{F}$ multiplets 
\begin{equation}
(\overline{5}+10,\overline{3})_{L}+(24,3)_{L}+2(1,3)_{R}+(1,\overline{6})_{R}
\label{fm}
\end{equation}%
which automatically appear free from both the $SU(5)$ and $SU(3)_{F}$
anomalies. They contain just three conventional families of quarks and
leptons plus massive multiplets located on the family symmetry scale $M_{F}$%
. In order to sufficiently suppress all flavor-changing transitions, which
would induce the family gauge boson exchanges, this scale should be at least
of the order $10^{5\div 6}$ $GeV$, though in principle it could be as large
as the $SU(5)$ GUT scale. In the latter case, some of the heavy states in (%
\ref{fm}) could be considered as candidates for the superheavy right-handed
neutrinos. One can argue that the physical composite multiplets (\ref{fm})
appear not only for the triple metacolor, $n=3$, but in general case as
well. Indeed, using the remark concerning the generalized AM condition (\ref%
{np}) and properly extending the left-handed multiplets in (\ref{216'}) and
the right-handed multiplets in (\ref{45}) by the new $n-3$ fundamental
composite octets $[(5,1)+(1,3)]_{L,R}$ to have anomaly matching for any
number $n$ of metacolors, one comes after pairing of the identical
multiplets to the same physical remnant (\ref{fm}) as in the triple
metacolor case.

It is important to note that the tiny radius of compositeness for universal
preons composing both quarks and leptons makes it impossible to directly
observe their composite nature \cite{ans}. Indeed, one can readily see that
the quark pair $u+d$ contains the same preons as the antiquark-antilepton
pair $\overline{u}+e^{+}$ that will lead to the process%
\begin{equation}
u+d\rightarrow \overline{u}+e^{+}
\end{equation}%
and consequently to the proton decay $p\rightarrow \pi ^{0}+e^{+}$ just due
to a simple rearrangement of preons inside the proton. To prevent this one
should take the compositeness scale $\Lambda _{MC}$ of the order of the
scale of the $SU(5)$\ GUT or even larger, $\Lambda _{MC}\gtrsim
M_{GUT}\approx 2\cdot 10^{16}$ GeV, and, respectively, $R_{MC}\leq 5\cdot
10^{-31}$ $sm$.

This limit on the radius of compositeness may in turn cause limits on the
composite fermions masses appearing as a result of the quantum gravitational
transitions of the identical states in the left-handed multiplets (\ref{216}%
) and right-handed multiplets (\ref{45}), 
\begin{equation}
(45,1)_{L,R}+(5,8+1)_{L,R}+(1,3)_{L,R}\text{ ,}  \label{fr}
\end{equation}%
into each other. From dimensional arguments related to a general structure
of the composites proposed above (\ref{wf}), these masses could be of the
order $(\Lambda _{MC}/M_{Pl})^{5}\Lambda _{MC}$ (that corresponds in fact to
the 6-fermion interaction of the left-handed and right-handed preons) and,
in fact, are very sensitive to the confinement\ scale $\Lambda _{MC}$.
Actually, for the metacolor scales, $M_{GUT}\leq \Lambda _{MC}\leq M_{Pl}$,
the heavy fermion masses may be located at the scales from $O(1)$ $TeV$ up
to the Planck mass scale. Therefore, the heavy composite states may be of
direct observation interest if they are located near the low limit, or
otherwise they will populate the $SU(5)$ GUT desert. Interestingly, the
screened preon states (\ref{wf2})%
\begin{equation}
(5,1)_{L,R}+(1,3)_{L,R}  \label{fr2}
\end{equation}%
acquire much heavier masses when being pairing with each other. Again, from
the dimensional arguments one may conclude that these masses has a natural
order $(\Lambda _{MC}/M_{Pl})\Lambda _{MC}$ that is significantly larger
than masses of the 3-preon states (\ref{wf}, \ref{wf1}).

Note that some of the heavy states (\ref{fr}) can mix with ordinary quarks
and leptons given by the multiplet $(\overline{5}+10,\overline{3})_{L}$ in (%
\ref{fm}). Particularly, there could be the large mixing term of the part $(%
\overline{5},\overline{3})_{L}$ containing the lepton doublet and down
antiquarks with the multiplet $(5,8+1)_{R}$ in (\ref{fr}). This term has a
form%
\begin{equation}
(\overline{5},\overline{3})_{L}(5,8+1)_{R}(1,3)  \label{53}
\end{equation}%
where $(1,3)$ stands for some pure "horizontal" scalar field being a triplet
of the family symmetry $SU(3)_{F}$. Actually, this mixing is related again
to the 6-fermion gravitational interaction of the left-handed and
right-handed preons, thus leading to the nondiagonal masses of the order $%
(\Lambda _{MC}/M_{Pl})^{5}M_{F}$. Thereby, in order not to significantly
disturb the masses of quarks and leptons in (\ref{fm}) one has to generally
propose $M_{F}\ll \Lambda _{MC}$. This is readily satisfied even for high
family scales, namely, in the case when the scale $M_{F}$ is taken near the
grand unification scale $M_{GUT}$, while the scale $\Lambda _{MC}$ near the
Planck scale $M_{Pl}$. The more liberal limitations appears when that part $(%
\overline{5},\overline{3})_{L}$ mixes with the screen preon states $%
(5,1)_{R} $ in (\ref{fr2}) due to the same scalar triplet $(1,3)$ of the $%
SU(3)_{F}$. Now, this mixing caused by the 4-fermion interaction leads to
the nondiagonal mass of the order $(\Lambda _{MC}/M_{Pl})^{2}M_{F}$ that may
be naturally much lesser than diagonal mass $(\Lambda _{MC}/M_{Pl})\Lambda
_{MC} $ derived above for the screened preon state. Nevertheless, depending
on real values of the scales $\Lambda _{MC}$ and $M_{F}$ there could be
expected some violation of unitarity in the conventional $3\times 3$ mass
matrices of leptons and down quarks which may be of a special interest for
observations. Other mixings of quarks and leptons with heavy states (\ref{fr}%
) and (\ref{fr2}) will necessarily include an ordinary Higgs quintet of the
grand unified $SU(5)$ (or a doublet of the SM) and, therefore, are
negligibly small.

To conclude, our preon model predicts three types of states which are: (1)
the three families of ordinary quarks and leptons $(\overline{5}+10,%
\overline{3})_{L}$ in (\ref{fm}) with masses at the electroweak scale, (2)
the heavy chiral multiplets $(24,3)_{L}+2(1,3)_{R}+(1,\overline{6})_{R}$ (%
\ref{fm}) with the Majorana type masses at the family scale $M_{F}=10^{6\div
16}$ $GeV$ and (3) the heavy paired multiplets (\ref{fr}) with masses in the
interval $10^{3\div 19}$ $GeV$ which are related to the gravitational
transition amplitudes of the $L$-preon composites into the $R$-preon ones.
However, the most important prediction of the left-right preon model\
considered here is, indeed, an existence of the local chiral family (or
horizontal) symmetry $SU(3)_{F}$ for quark-lepton generations which is
briefly presented below.

\section{The chiral family symmetry SU(3)$_{F}$}

The flavor mixing of quarks and leptons is certainly one of the major
problems that presently confront particle physics. Many attempts have been
made to interpret the pattern of this mixing in terms of various family
symmetries - discrete or continuous, global or local. Among them, the chiral
family symmetry $SU(3)_{F}$ derived first in the similar preon framework 
\cite{jp} and developed then by its own by many authors [13-22] seems most
promising. As was shown, the spontaneous breaking of this symmetry gives
some guidance to the observed hierarchy between elements of the quark-lepton
mass matrices, on the one hand, and to presence of texture zeros in them, on
the other, that leads to relationships between the mass and mixing
parameters. In the framework of the supersymmetric Standard Model, it leads,
at the same time, to an almost uniform mass spectrum for the superpartners,
with a high degree of flavor conservation, that makes its existence even
more significant in the SUSY case.

Generically, the chiral family symmetry $SU(3)_{F}$ possesses four basically
attractive features:

(i) It provides a natural explanation of the number three of observed
quark-lepton families, correlated with three species of massless or light ($%
m_{\nu }<M_{Z}/2$) neutrinos contributing to the invisible $Z$ boson partial
decay width;

(ii) Its local nature conforms with the other local symmetries of the
Standard Model, such as the weak isospin symmetry $SU(2)_{w}$ or color
symmetry $SU(3)_{c}$, thus leading to the family-unified SM with a total
symmetry $SM\times SU(3)_{F}$;

(iii) Its chiral nature, according to which both left-handed and
right-handed fermions are proposed to be fundamental triplets of the $%
SU(3)_{F}$, provides the hierarchical mass spectrum of quark-lepton families
as a result of a spontaneous symmetry breaking at some high scale $M_{F}$
which could in principle located in the area from $10^{5\div 6}$ GeV (to
properly suppress the flavor-changing processes) up to the grand unification
scale $M_{GUT}$ and even higher. Actually, any family symmetry should be
completely broken in order to conform with reality at lower energies. This
symmetry should be chiral, rather than a vectorlike, since a vectorlike
symmetry would not forbid the invariant mass, thus leading in general to
degenerate rather than hierarchical mass spectra. Interestingly, both known
examples of local vectorlike symmetries, electromagnetic $U(1)_{EM}$ and
color $SU(3)_{C}$, appear to be exact symmetries, while all chiral
symmetries including conventional grand unifications \cite{moh} $SU(5)$, $%
SO(10)$ and $E(6)$ (where fermions and antifermions lie in the same
irreducible representations) appear broken;

(iv) Thereby, due to its chiral structure, the $SU(3)_{F}$ admits a natural
unification with all known GUTs in a direct product form, both in an
ordinary and supersymmetric framework, thus leading to the family-unified
GUTs, $GUT\times SU(3)_{F}$, beyond the Standard Model.

So, if one takes these naturality criteria seriously, all the candidates for
flavor symmetry can be excluded except for local chiral $SU(3)_{F}$
symmetry. Indeed, the $U(1)$ family symmetry does not satisfy the criterion
(i) and is in fact applicable to any number of quark-lepton families. Also,
the $SU(2)$ family symmetry can contain, besides two light families treated
as its doublets, any number of additional (singlets or new doublets of $%
SU(2) $) families. All the global non-Abelian symmetries are excluded by
criterion (ii), while the vectorlike symmetries are excluded by the last
criteria (iii) and (iv).

Among applications of the $SU(3)_{F}$ symmetry, the most interesting ones
are the description of the quark and lepton masses and mixings in the
Standard Model and GUTs \cite{su3}, neutrino masses and oscillations \cite%
{ber1} and rare processes \cite{jon0} including their astrophysical
consequences \cite{m}. Remarkably, the $SU(3)_{F}$ invariant Yukawa coupling
are always accompanied by an accidental global chiral $U(1)$ symmetry, which
can be identified with the Peccei-Quinn symmetry \cite{peccei} provided it
is not explicitly broken in the Higgs sector, thus giving a solution to the
strong CP problem \cite{q84}. In the SUSY context \cite{q86}, the $SU(3)_{F}$
model leads to a special relation between (s)fermion masses and the soft
SUSY breaking terms at the GUT scale in a way that all the dangerous
flavor-changing processes are naturally suppressed. The special sector of
applications is related to a new type of topological defects - flavored
cosmic strings and monopoles appearing during the spontaneous violation of
the $SU(3)_{F}$ which may be considered as possible candidates for the cold
dark matter in the Universe \cite{def}.

Let us note in conclusion that if the family symmetry $SU(3)_{F}$ arises
from the preon model proposed above one can expect that in the emerged $%
SU(5)\times SU(3)_{F}$ GUT the gauge coupling constants $g_{5}$ and $g_{3F}$
should be equal at the $SU(8)_{MF}$ unification scale. The study of flavor
changing processes $\mu \rightarrow e+\gamma $, $D^{0}-\overline{D}^{0}$, $%
B^{0}-\overline{B}^{0}$ and others caused by the $SU(3)_{F}$ gauge boson
exchanges could in principle show whether the family symmetry has an origin
in the preon model or it is, rather, independently postulated. However, the
most crucial difference between these two cases is related to the existence
in the preon model of some heavy $SU(5)\times SU(3)_{F}$ multiplets located
at scales from $O(1)$ $TeV$ up to the Planck mass. If they are relatively
light, they may be of direct observational interest by them own. If they are
heavy, they still strongly affect the quark-lepton mass matrices due to
their large mixings with the down quarks and leptons, as was shown in (\ref%
{53}). Remarkably, even if the family symmetry $SU(3)_{F}$ is taken at the
GUT scale the difference between these cases is still left. Indeed, now all
flavor-changing transitions due to the family gauge boson exchange will be
extremely suppressed, while for the independently introduced family symmetry
these transitions may significantly contribute into rates of the nondiagonal
processes. Moreover, for the high scale family symmetry one has some natural
candidates for massive right-handed neutrinos in terms of the extra heavy
states given in (\ref{fm}).

\section{Conclusion and outlook}

We have shown that, apart from somewhat inspirational religious and
philosophical aspects ensured by the Eightfold Way, the $SU(8)$ symmetry as
a basic internal symmetry of the physical world is indeed advocated by preon
model for composite quarks and leptons.

In fact, many preon models have been discussed and considered in the past
(some significant references can be found in \cite{moh, ds}), though they
were not turned out to be too successful and attractive, especially compared
with other theory developments, like as supersymmetry and supergravity,
appeared at almost the same time. However, there is still left a serious
problem in particle physics with classification of all observed quark-lepton
families. As in the hadron spectroscopy case, this may motivate us to
continue seeking a solution in some subparticle or preon models for quarks
and leptons, rather than in the less definitive extra dimension or
superstring theories.

Let us briefly recall the main results presented here. We have started with
the $L$-$R$ symmetric preon model and found that an admissible metaflavor
symmetry $SU(8)_{MF}$ appears as a solution to the 't Hooft's anomaly
matching condition providing preservation of the accompanying chiral
symmetry $SU(8)_{L}\times SU(8)_{R}$ at all scales involved. In contrast to
a common point of view, we require that states composed from the left-handed
and right-handed preons with their own metacolors, $SO(3)_{MC}^{L}\times
SO(3)_{MC}^{R}$, have to satisfy AM condition separately, though their
triangle anomalies may compensate each other. The point is, however, such an
emerged $L$-$R$ symmetric $SU(8)_{MF}$ theory, though is chiral with respect
to preons, certainly appears vectorlike for the identical $L$-preon and $R$%
-preon composite multiplets involved. As a consequence, while preons are
left massless being protected by their own metacolors, all $L$-preon and $R$%
-preon composites being metacolor singlets will pair up and, therefore,
acquire superheavy Dirac masses. It is rather clear that such a theory is
meaningless unless the $L$-$R$ symmetry is partially broken that seems to be
a crucial point in our model. In this connection, some natural mechanisms
for spontaneous $L$-$R$ symmetry breaking have been proposed according to
which some $R$-preons, in contrast to $L$-preons, may be condensed or such
asymmetry may be caused by the properly arranged scalar field potential. As
result, an initially emerged vectorlike $SU(8)_{MF}$ theory reduces down to
the conventional $SU(5)$ GUT with an extra local family symmetry $SU(3)_{F}$
and three standard generations of quarks and leptons. Though the tiny radius
of compositeness for universal preons composing both quarks and leptons
makes it impossible to immediately confirm their composite nature, a few
special $SU(5)\times SU(3)_{F}$ multiplets of extra composite fermions
located at the scales from $O(1)$ $TeV$ up to the Planck mass scale
predicted by the theory may appear of actual experimental interest. Some of
them can be directly observed, the others populate the $SU(5)$ GUT desert.
Due to their mixing with ordinary quark-lepton families there may emerge a
marked violation of unitarity in the mass matrices for leptons and down
quarks depending on the interplay between the compositeness scale and scale
of the family symmetry $SU(3)_{F}$.

For the reasons of simplicity, we have not considered here boson composites
which could appear as the effective scalar or vector fields in the theory.
Generally, they will become very heavy (with masses of the order of the
compositeness scale $\Lambda _{MC}$) unless their masses are specially
protected by the low-scale supersymmetry. The point is, however, that some
massless composite vector fields could nonetheless appear in a theory as the
Goldstone bosons related to spontaneous violation of Lorentz invariance
through the multi-preon interactions similar to those (\ref{4f}) given in
the section 5. In principle, one could start with a global metaflavor
symmetry $SU(N)_{MF}$ which is then converted into the local one through the
contact multi-preon interactions \cite{jb} or some nonlinear constraint put
on the preon currents (see in this connection \cite{nam} and the later works 
\cite{cfn}). If so, the quarks and leptons, on the one hand, and the gauge
fields (photons, weak bosons, gluons etc.), on the other, could be composed
at the same order distances determined by the preon confinement scale $%
\Lambda _{MC}$. In other words, there may be a lower limit to the division
of matter beyond which one can not go. Indeed, a conventional division of
matter from atoms to quarks is naturally related to the fact that matter is
successively divided, whereas the mediator gauge fields are left intact.
However, situation may be drastically changed if these spontaneously
emerging gauge fields become composite as well. We will address this and
other related questions elsewhere.

\section*{ Acknowledgments}

I am grateful to many people who had significantly contributed to the ideas
presented here during the years when they were developed, especially to A.A.
Anselm, V.N. Gribov, S.G. Matinyan and V.I. Ogievetsky, as well as to my
collaborators Z.G. Berezhiani, O.V. Kancheli and K.A. Ter-Martirosyan. I
would also like to thank the organizers and participants of the 20th
International Workshop "What Comes Beyond the Standard Model?" (9-17 July
2017, Bled, Slovenia) M.Yu. Khlopov, N.S. Manko\v{c} Bor\v{s}tnik, H.B.F.
Nielsen and K.V. Stepanyantz for a warm hospitality and interesting
discussions.


\begin{thebibliography}{99}
\bibitem{ew} M. Gell-Mann and Y. Ne'eman, \textit{The Eightfold Way} (W.A.
Benjamin, New York, 1964).

\bibitem{nep} https://en.wikipedia.org/wiki/Noble\_Eightfold\_Path.

\bibitem{t} G. 't Hooft, in \textit{Recent Developments in Gauge Theories},
edited by G.'t Hooft et al (Plenum, New-York, 1980).

\bibitem{jp} J.L. Chkareuli, JETP Lett. 32 (1980) 671;

J.L. Chkareuli , \textit{Composite Quarks And Leptons: From Su(5) To Su(8)
Symmetry}, In Proc. of the Int. Seminar \textit{QUARKS-82}, pp 149-156,
edited by A.L. Kataev (INR, Moscow, 1982).

\bibitem{moh} R.N. Mohapatra, \textit{Unification and Supersymmetry}
(Springer-Verlag, New-York, 2003).

\bibitem{ds} I.A. D'Souza and C.S. Kalman, \textit{Preons: Models of
Leptons, Quarks and Gauge Bosons as Composite Objects }(World Scientific,
Singapore, 1992);

H. Terazawa, Y. Chikashige and K. Akama, Phys. Rev. D 15 (1977) 480 ; H.
Terazawa, ibid. 22 (1980) 184.

\bibitem{th} G. 't Hooft, Phys. Rev. Letters 37 (1976) 8.

\bibitem{bar} R. Barbieri, L. Maiani and R. Petronzio, \ Phys. Lett. B 96
(1980) 63.

\bibitem{wil} K.G. Wilson, Phys. Rev. 10 (1974) 2245;

M. Creutz, Phys. Rev. Lett. 43 (1979) 553.

\bibitem{io} Y. Nambu and G. Jona-Lasinio, Phys. Rev. 122 (1961) 345.

\bibitem{jb} J.D.~Bjorken, Ann. Phys. (N.Y.) \ 24\textbf{\ }(1963) 174;

Per Kraus and E.T. Tomboulis, Phys. Rev. D 66 (2002) 045015.

\bibitem{ans} A.A. Anselm, Sov. Phys. JETP 53 (1981) 23.

\bibitem{su3} Z.G. Berezhiani and J.L. Chkareuli, Sov. J. Nucl. Phys. 37
(1983) 618;

Z.G. Berezhiani, Phys. Lett. B129 (1983) 99; ibid B150 (1985) 177;

Z. Berezhiani and M. Yu.Khlopov, Sov. J. Nucl. Phys. 51(1990) 935;

T. Appelquist, Y. Bai and M. Piai, Phys. Lett. B 637 (2006) 245 .

\bibitem{wf} F. Wilczek, AIP Conf. Proc. 96 (1983) 313.

\bibitem{ber1} Z. G. Berezhiani, J. L. Chkareuli, JETP Lett. 37 (1983) 338;

Z.G. Berezhiani and J.L. Chkareuli, Sov. Phys. Usp. 28 (1985) 104;

T. Appelquist, Y. Bai and M. Piai, Phys. Rev. D 74 (2006) 076001 .

\bibitem{jon0} J.L. Chkareuli, Phys. Lett. B246 (1990) 498; ibid B 300
(1993) 361.

\bibitem{q84} Z.G. Berezhiani and J.L. Chkareuli , \textit{Horizontal
Unification Of Quarks And Leptons}, In Proc. of the Int. Seminar \textit{%
QUARKS-84}, vol. 1, pp 110-121, edited by A.N. Tavkhelidze et al (INR,
Moscow, 1984);

Z. Berezhiani and M. Yu.Khlopov, Sov. J. Nucl. Phys. 51(1990) 739;

T. Appelquist, Y. Bai and M. Piai, Rev. D 75(2007) 073005.

\bibitem{peccei} R.D. Peccei and H.R. Quinn, Phys. Rev. D16 (1977) 1891.

\bibitem{q86} Z.G. Berezhiani, J.L. Chkareuli, G.R. Dvali and M.R. Jibuti, 
\textit{Supersymmetry and generations of quarks and leptons}, In Proc. of
the Int. Seminar \textit{QUARKS-86}, pp 209-223, edited by A.N. Tavkhelidze
et al (INR, Moscow, 1986);

Z. Berezhiani, Phys. Lett. B 417 (1998) 287;

S.F. King and G.G. Ross, Phys. Lett. B 520 (2001) 243;

J.L. Chkareuli, C.D. Froggatt and H.B. Nielsen, Nucl. Phys. B 626 (2002) 307.

\bibitem{def} J.L. Chkareuli, Phys. Lett. B272 (1991) 207;

G. Dvali and G. Senjanovic, Phys. Rev. Lett. 72 (1994) 9;

D. Spergel and U.-Li Pen, Astrophys. J. 491 (1997) L67;

S.M. Carroll and M. Trodden , Phys. Rev. D57 (1998) 5189.

\bibitem{r} P. Ramond, hep-ph/9809459.

\bibitem{m} M. Yu. Khlopov, \textit{Cosmoparticle Physics} (World
Scientific, Singapore, 1999).

\bibitem{nam} Y. Nambu, Progr. Theor. Phys. Suppl. E 68 (1968) 190.

\bibitem{cfn} J.L. Chkareuli, C.D. Froggatt, J.G. Jejelava and H.B. Nielsen,
Nucl. Phys. B 796 (2008) 211; J.L. Chkareuli, C.D. Froggatt and H.B.
Nielsen, Nucl. Phys. B 848 (2011) 498.
\end{thebibliography}
\end{document}